\documentclass[pre,twocolumn,showpacs]{revtex4}
\usepackage{epsfig}
\begin{document}

\author{A. Rebenshtok, E. Barkai}
\affiliation{Department of Physics,
Bar Ilan University, Ramat-Gan 52900 Israel}

\title{Weakly non-ergodic Statistical Physics}


\begin{abstract}
 
 For weakly non ergodic systems,
the probability
density function of a time average observable $\overline{{\cal O}}$ is 
$ f_{\alpha} \left( \overline{{\cal O}} \right) =
- { 1 \over \pi} \lim_{\epsilon\to 0} \mbox{Im}
{ \sum_{i=1} ^L p^{{\rm eq}} _i \left( \overline{{\cal O}} - {\cal O}_i + i \epsilon\right)^{\alpha -1} \over
 \sum_{i=1} ^L p^{{\rm eq}} _i \left( \overline{{\cal O}} - {\cal O}_i + i \epsilon\right)^\alpha}
$
where ${\cal O}_i$ is the value of the observable when the system
is in state $i=1,\cdots L$. $p^{{\rm eq}}_i$ is the 
probability that a member
of an ensemble of systems occupies state $i$ in equilibrium. 
For a particle
undergoing a fractional diffusion process in a binding force field,
with thermal detailed
balance conditions,
 $p^{{\rm eq}}_i$ is Boltzmann's canonical probability. Within
the unbiased sub-diffusive continuous time random walk model, the exponent
$0<\alpha<1$ is the anomalous diffusion exponent 
$\langle x^2 \rangle \sim t^\alpha$ found for free boundary conditions.
 When $\alpha \to 1$ ergodic statistical mechanics is recovered
$ \lim_{\alpha \to 1} f_{\alpha} (\overline{{\cal O}}) = \delta\left( \overline{{\cal O}} - \langle {\cal O} \rangle \right)$.
We briefly discuss possible physical applications
in single particle experiments.  

\end{abstract}
\maketitle

\section{Introduction}

 An ensemble of non-interacting one dimensional
Brownian particles in the presence of a binding potential
field $V(x)$ reach a thermal equilibrium described by
Boltzmann's canonical law $p^{{\rm eq}}(x)= \exp[ - V(x)/T]/Z$
where $T$ is the temperature (units $k_b=1$) and $Z$ is the
normalizing partition function. With this law we may calculate
ensemble averages for example $\langle x \rangle = \int_{-\infty} ^\infty
x p^{{\rm eq}} (x) {\rm d} x$. On the other hand from  the trajectory
of a single particle $x(t)$ we
may construct the time average $\overline{x}= \int_0 ^t x(t') {\rm d} t' / t$.
For ergodic motion the time and ensemble averages are identical
in the limit of long measurement times $t \to \infty$. What is
the Physical meaning of a long measurement time? 
Brownian dynamics in a finite interval, is characterized by a 
finite relaxation time which is the time scale on which 
particles reach thermal equilibrium. For the simplest case of
Brownian motion between two reflecting walls, with a system of size
$l$, dimensional analysis gives a relaxation time of the
order $l^2 / D$ where $D$ is the diffusion constant. A
second time scale, the average time between jump events
$\langle \tau \rangle$ is more microscopical.
The latter  is related to $D$ with the Einstein relation
\cite{vanKampen}
%
%\begin{equation}
$D = \langle (\Delta x)^2 \rangle /  2 \langle \tau \rangle$
%\label{eq0}
%\end{equation}
%
where $\langle( \Delta x )^2 \rangle$ is the variance of jump lengths.
Ergodicity is almost trivial when the measurement time
is much longer than these two time scales. 
For example for a Brownian motion in a Harmonic field.

 On the other hand anomalous diffusion and transport, is characterized
in many cases by a diverging relaxation time and a diverging microscopical
time scale $\langle \tau \rangle$. For example unbiased
sub-diffusion is characterized by a mean square displacement
$\langle x^2 \rangle \propto t^\alpha $ and $0<\alpha < 1$. The reader 
immediately realizes that the diffusion constant is zero,
in the sense that $\lim_{t \to \infty} \langle x^2 \rangle / t = 0$,
hence the mentioned relaxation time $l^2 /D$ is infinite 
even when the system size $l$ is finite. 
Indeed according to the continuous time random walk (CTRW) model 
\cite{SM,BouchaudREV,KSZ,Metzler,Klages},
anomalous sub-diffusion is found when waiting times between jumps diverge
$\langle \tau \rangle \to \infty$, which is related to the Scher-Montroll
power law waiting time probability density function (PDF)
$\psi(\tau) \propto \tau^{ - (1 + \alpha)}$ \cite{SM}.
For such scale
free anomalous diffusion the relaxation time and the averaged
sojourn time $\langle \tau \rangle$ are infinite, and 
ergodicity is broken weakly. 

 Strong ergodicity breaking is found when a system is divided into
inaccessible regions of its phase space. Namely a particle or a system
starting in one region cannot explore all other regions due to 
some non-passable barrier  (e.g. in the
micro-canonical case, by regions we mean sections on the
constant energy surface). 
Bouchaud  \cite{WEB} introduced the profound concept of
weak ergodicity breaking in the context of glass dynamics,
which in turn is  related to 
infinite ergodic theory \cite{Aaronson} investigated by Mathematicians  
using a dynamical approach. In weak ergodicity breaking the phase
space is not broken into inaccessible regions. Instead
due to the power law sticking times the dynamics is non-stationary
and non-ergodic. 
Since the ergodic hypothesis is the pillar on which
statistical mechanics is built, but at the same time also 
long tailed power law distributions of 
trapping times are very common in the description of
Physical behaviors \cite{SM,BouchaudREV,KSZ,Metzler,Klages},
it is natural to investigate the
non-ergodic properties of systems whose stochastic dynamics are
governed by such anomalous statistics. 
Previously weak ergodicity
breaking was investigated for: blinking quantum dots 
\cite{Dahan,Marg,Marg1,Marg3}, L\'evy walks \cite{Marg2}
occupation time statistics
of the CTRW model \cite{Bel1,Bel2}, fractional Fokker-Planck equation
\cite{BarkaiJSP},
deterministic one dimensional maps \cite{Thaler,Euro,Japan},   
 numerical
simulations of fractional transport in a washboard potential
\cite{Goy} and
in vivo gene regulation by DNA-binding proteins \cite{Lomholt}.
Recently a relation between statistics of
weak ergodicity breaking and statistics of non-self averaging
in models of quenched disorder was found \cite{Stas}. 
Hence it is timely to present a general statistical mechanical 
framework for weak-ergodicity
breaking. 

In this manuscript we investigate 
the distribution of time averaged
observables for weak ergodicity breaking.
We explore the relations between ensemble averages and
fluctuations of time averages.   
And investigate the transition from the localization limit $\alpha \to 0$
to the usual ergodic behavior
found for $\alpha \to 1$. Specific examples for the distribution
of $\overline{x}$  for a particle undergoing a sub-diffusive
CTRW in a potential $V(x)$ are worked out is detail. 
In the second part of the paper we derive our main results 
using a generalized CTRW approach. We investigate models with a single
waiting time PDF and more general models
with several types of such PDFs.   
A brief summary of some of our results was recently published
in \cite{RebPRL}. 

 The theory of weak-ergodicity breaking is mathematically related
to the arcsine distribution \cite{Feller,Redner,Godreche},
and to its extensions \cite{Lamp,Barlow}.
Consider a
normal Brownian motion with free boundary conditions
in one dimension $\dot{x}(t) = \eta(t)$ where
$\eta(t)$ is Gaussian white noise. The time $t^{+}$ spent 
by the particle in $x>0$ is called an occupation time. Naive
expectation is that the single particle will occupy $x>0$ for
half of the measurement time $t$, when the latter is long. Instead
the PDF of the occupation fraction is  
\begin{equation}
f\left( {t^{+} \over t} \right) = { 1 \over \pi \sqrt{ \left( t^{+}/t \right) \left( 1 - t^{+}/t\right) } }.
\label{eqARC}
\end{equation}
This arcsine law
 is related to the well known PDF of first-passage times
from  $x$ to the origin, for simple Brownian motion.
The latter PDF decays $t^{-3/2}$ for long first-passage times \cite{Redner},
and the averaged return time is infinite. Roughly speaking,
during the dynamics of
the particle, it  will usually occupy either the domain $x>0$
or the domain $x<0$ for a duration which is  of the order
of the measurement time. Thus the arcsine PDF Eq. (\ref{eqARC})
has a $U$ shape.   
This behavior is found because the Brownian motion was assumed to be
unbounded. If we add reflecting walls on $x=l/2$ and
$x=-l/2$  the dynamics will be ergodic and in the long time
limit the particle 
spends half of the time in $(0,l/2)$ and the other half in $(-l/2,0)$. 
The theory of weak ergodicity breaking, presented in this manuscript, is
mathematically related to the arcsine law and is based on L\'evy
statistics. 

 This paper is organized as follows. In Sec. \ref{SecDIS} distribution
of time averaged observables for weakly non-ergodic systems is
presented using general arguments not specific to a model. 
Properties of this distribution
are investigated in Sec. 
\ref{SecStat} and in Sec. 
\ref{SecX} the example of the distribution of $\overline{x}$ for a particle
undergoing fractional dynamics in a binding potential 
is worked out in detail.
In Sec. 
\ref{SecCTRW} we derive our main result using a CTRW approach, thus further
justifying
assumptions made in Sec. \ref{SecDIS}. Numerical simulations of 
$\overline{x}$ obtained from the
CTRW process are compared with analytical theory in sub-Sec.  
\ref{subSnum}.

\section{Distribution of Time Averaged Observables}
\label{SecDIS}

 We consider a system with $L$ states and label them with an index $i=1,...L$.
 A time average of a  Physical observable
${\cal O}$ is made.  If the system is in state $i$ the Physical
observable  attains the value
${\cal O}_i $.  
The time the system spends in state $i$ is $t_i$ and is
called a residence time or an occupation time.
The
time average of the Physical observable is 
\begin{equation}
\overline{{\cal O}} = {\sum_{i=1} ^L t_i {\cal O}_i \over \sum_{i=1} ^L t_i}
\label{wq01}
\end{equation}
and
$\mbox{min}\{ {\cal O}_i \} \le \overline{{\cal O}} \le \mbox{max}\{ {\cal O}_i \}$.
As mentioned in the introduction
many physical systems, in their stochastic or deterministic dynamics, 
are known to be characterized
by power law sojourn times in the states of the system
\cite{SM,BouchaudREV,KSZ,Metzler,Klages}. 
We assume that the occupation time $t_i$ is a sum
of many such sojourn times.  If the
state $i$ is visited many times, and the sojourn times are independent
identically distributed random variables,  L\'evy's limit theorem
will describe the statistics of the residence 
times $t_i$ in the limit
of long measurement time.
Hence we argue that the PDF of $t_i$
is a one sided L\'evy PDF
$l_{\alpha, p^{{\rm eq}} _i} (t_i)$
whose
Laplace transform is 
\begin{equation}
\int_0 ^\infty l_{\alpha,p^{{\rm eq}}_i } (t_i) \exp( - u t_i){\rm d} t_i
= \exp\left( - p^{{\rm eq}}_i u^\alpha\right)
\label{eq02}
\end{equation}  
and $0<\alpha\le 1$ \cite{remark}. 
Later we find that $p^{{\rm eq}}_i$ is the probability that a member
of an ensemble of systems occupies state $i$ in equilibrium.
For the CTRW with thermal detailed balance conditions $p^{{\rm eq}}_i$
is Boltzmann's probability of finding the system in state
$i$, $p^{{\rm eq}}_i = \exp( - E_i /T)/Z$ as we will show later. 

 The following  generating function \cite{Godreche} is a useful tool 
\begin{equation}
\hat{f}_{\alpha} \left( \xi \right) = 
\langle {1 \over 1 + \xi \overline{{\cal O}} } \rangle = \sum_{k=0} ^\infty \left( - 1 \right)^k \langle \overline{{\cal O}}^k \rangle \xi^k. 
\label{eq03}
\end{equation}
Our main aim is to find the PDF of the time averaged observable
\begin{widetext}
\begin{equation}
f_{\alpha} \left( \overline{{\cal O}} \right) = \langle \delta\left(
\overline{{\cal O}} - { \sum_{i=1} ^L {\cal O}_i t_i \over \sum_{i=1} ^L t_i } \right) \rangle=  
 - {1 \over \pi} \lim_{\epsilon \to 0}\mbox{Im} \langle {1 \over \overline{{\cal O}} + i \epsilon - { \sum_{i=1} ^L {\cal O}_i t_i \over \sum_{i=1} ^L t_i } } \rangle =
 - {1 \over \pi} \lim_{\epsilon \to 0} \mbox{Im} {1 \over \overline{{\cal O}} + i \epsilon}  \langle {1 \over 1-  {1 \over \overline{{\cal O}} + i \epsilon} {\sum_{i=1} ^L {\cal O}_i t_i \over \sum_{i=1} ^L t_i } } \rangle, 
\label{eq000}
\end{equation}
using Eq. 
(\ref{eq03})
\begin{equation}
f_{\alpha} \left( \overline{{\cal O}} \right) = 
- {1 \over \pi} \lim_{\epsilon \to 0} \mbox{Im} {1 \over \overline{{\cal O}} +i \epsilon } 
\hat{f}_{\alpha} \left( - {1 \over \overline{{\cal O}} + i \epsilon} \right)
%=
%- {1 \over \pi} \lim_{\epsilon \to 0} \left[
% { \overline{{\cal O }}  \over \overline{{\cal O}}^2 + \epsilon^2} \mbox{Im} 
%\hat{f}_{\alpha} \left( - {1 \over \overline{{\cal O}} + i \epsilon} \right) -
% { \epsilon \over \overline{{\cal O}}^2 + \epsilon^2} \mbox{Re} 
%\hat{f}_{\alpha} \left( - {1 \over \overline{{\cal O}} + i \epsilon} \right) 
%\right]
\label{eq04a}
\end{equation}
%
%hence \cite{Godreche}
%
%\begin{equation}
%f_{\alpha} \left( \overline{{\cal O}} \right) = 
%- {1 \over \pi} \lim_{\epsilon \to 0} \mbox{Im} {1 \over \overline{{\cal O}} } 
%\hat{f}_{\alpha} \left( - {1 \over \overline{{\cal O}} + i \epsilon} \right)
%\label{eq04}
%\end{equation}
%
%and for  $f_{\alpha}(\overline{O} =0)$ we use 
%$\lim_{\overline{O} \rightarrow 0} f_{\alpha}(\overline{O})$.
We now find the generating function   
Eq. (\ref{eq03}) and invert it using Eq. (\ref{eq04a}) 
to obtain $f_\alpha\left( \overline{{\cal O}} \right)$.

 The generating function is rewritten
$$\hat{f}_\alpha \left( \xi \right) = \langle \int_0 ^\infty {\rm d} s e^{ - \left( 1 + \xi { \sum_{i = 1} ^L {\cal O}_i t_i \over \sum_{i = 1} ^L t_i } \right) s }  \rangle = $$ 
\begin{equation}
\int_0 ^\infty {\rm d} s \int_0 ^\infty {\rm d} t \int_0 ^\infty {\rm d} t_1 l_{\alpha,p^{{\rm eq}} _1} \left( t _1 \right) \cdots 
\int_0 ^\infty {\rm d} t_L l_{\alpha,p^{{\rm eq}} _L} \left( t _L \right)  
\delta\left( t - \sum_{i=1} ^L t_i \right) 
e^{ - \left( 1 + \xi { \sum_{i = 1} ^L {\cal O}_i t_i \over  t } \right) s }. 
\label{eq05}
\end{equation}
Using a well known presentation for the delta function in Eq. (\ref{eq05}) 
\begin{equation}
\delta\left(t- \sum_{i =1} ^L t_i\right) = 
 {1 \over 2 \pi} \int_{- \infty} ^\infty {\rm d} k e^{ i k \left( t - \sum_{i =1} ^L t_i\right) }, 
\label{eq06}
\end{equation}
changing variables  
$k t = \tilde{k}$ 
and using Eq. 
(\ref{eq02}) we find
\begin{equation}
\hat{f}_\alpha \left( \xi \right) = \int_0 ^\infty {\rm d} t t^{-1}
\int_{-\infty} ^\infty {{\rm d} \tilde{k} \over 2 \pi} \int_0 ^\infty {\rm d} s \exp \left[ i \tilde{k} - s 
- \sum_{i=1} ^L p^{{\rm eq}}_i { \left( i \tilde{k} + {\cal O}_i \xi s \right)^\alpha \over t^\alpha } \right].
\label{eq07}
\end{equation}
We again change variables $k = \tilde{k}/t$ and $\tilde{s}=s/t$
and obtain
\begin{equation}
\hat{f}_\alpha \left( \xi \right) = \int_0 ^\infty {\rm d}t 
\int_{-\infty} ^\infty {{\rm d} k \over 2 \pi} \int_0 ^\infty {\rm d} \tilde{s} t \exp \left[ i k t - \tilde{s} t  
- \sum_{i=1} ^L p^{{\rm eq}}_i  \left( i k + {\cal O}_i \xi \tilde{s} \right)^\alpha \right].
\label{eq08}
\end{equation}
This equation is rewritten using a simple trick
$$ \hat{f}_\alpha \left( \xi \right) =
\int_0 ^\infty {\rm d} t \int_{\-\infty} ^\infty {{\rm d} k \over 2 \pi } 
\int_0 ^\infty {\rm d} \tilde{s} \left\{ 
- {{\rm d} \over {\rm d} \tilde{s} } \exp\left[
i k t - \tilde{s} t - \sum_{i=1} ^L p^{{\rm eq}}_i  \left( i k + {\cal O}_i \xi \tilde{s} \right)^\alpha \right] \right. 
$$
\begin{equation}
\left.
- \alpha \sum_{i=1} ^L p^{{\rm eq}}_i \left( i k + {\cal O}_i \xi \tilde{s} \right)^{\alpha-1} {\cal O}_i \xi \exp\left[ \left( i k - \tilde{s} \right) t
- \sum_{i=1} ^L p^{{\rm eq}}_i  \left( i k + {\cal O}_i \xi
\tilde{s} \right)^\alpha \right] \right\}.
\label{eq09}
\end{equation} 
\end{widetext}
Integration over $t$ gives a simple pole $1/(ik -\tilde{s})$, using
Cauchy integral formula to integrate over $k$, and then solving two trivial
integrals yields
\begin{equation}
\hat{f}_\alpha \left( \xi \right) =
{\sum _{i = 1} ^L p^{{\rm eq}}_i \left( 1 + {\cal O}_i \xi \right)^{\alpha-1} \over \sum_{i=1} ^L p^{{\rm eq}}_i \left( 1 + {\cal O}_i \xi \right)^\alpha }. 
\label{eq10}
\end{equation} 
Inverting Eq. (\ref{eq10}) using Eq. 
(\ref{eq04a}) we find 
\begin{equation}
f_{\alpha} \left( \overline{{\cal O}} \right) =
- { 1 \over \pi} \lim_{\epsilon\to 0} \mbox{Im}
{ \sum_{i=1} ^L p^{{\rm eq}} _i \left( \overline{{\cal O}} - {\cal O}_i + i \epsilon\right)^{\alpha -1} \over
 \sum_{i=1} ^L p^{{\rm eq}} _i \left( \overline{{\cal O}} - {\cal O}_i + i \epsilon\right)^\alpha}.
\label{eq11}
\end{equation}
This is a very general formula for the
distribution of time averaged observables for weakly non-ergodic
systems. As we show later, within the  CTRW model, $\alpha$ is
the anomalous diffusion exponent. 
 For 
$0<\alpha<1$ 
we use
$\lim_{\epsilon\to 0} \left( \overline{{\cal O}} - {\cal O}_i + i
\epsilon\right)^\alpha
= | \overline{{\cal O}}-{\cal O}_i |^\alpha e^{i \phi_i \alpha}$
where
$
 \phi_{i} = \left(
\begin{array}{c}
{\pi \textrm{ if } \overline{{\cal O}}<{\cal O}_i} \\
{0 \textrm{ if } \overline{{\cal O}} \ge {\cal O}_i} \\
\end{array}
\right)
$
and Eq.(\ref{eq11}) becomes
$$ f_{\alpha} \left( \overline{{\cal O}} \right) = $$
\begin{equation}
{ \sin \pi \alpha \over \pi}
{I^{<} _{\alpha - 1} \left( \overline{{\cal O}} \right) I^{\ge}
_{\alpha}\left( \overline{{\cal O}} \right) + I^{\ge} _{\alpha -1}
\left( \overline{{\cal O}} \right) I^{<}_{\alpha} \left(
\overline{{\cal O}} \right)\over
\left[ I^{\ge} _{\alpha } \left( \overline{{\cal O}} \right)\right]^2
+
\left[ I^{<} _{\alpha } \left( \overline{{\cal O}} \right)\right]^2
+
2 I^{\ge} _{\alpha } \left( \overline{{\cal O}} \right) I^{<} _{\alpha
} \left( \overline{{\cal O}} \right) \cos \pi \alpha },
\label{eq40Adi}
\end{equation}
with
\begin{equation}
I^{<} _{\alpha } \left( \overline{{\cal O}} \right)=
\sum_{\overline{{\cal O}} < {\cal O}_{i}} p^{{\rm eq}}_i |
\overline{{\cal O}} - {\cal O}_i|^{\alpha}
\label{eq38ADI}
\end{equation}
and
\begin{equation}
I^{\ge} _{\alpha } \left( \overline{{\cal O}} \right)=
\sum_{\overline{{\cal O}} \ge {\cal O}_{ i}} p^{{\rm eq}}_i |
\overline{{\cal O}} - {\cal O} _i|^{\alpha}.
\label{eq38ADIa}
\end{equation}
Notice the $L$ divergences of $f_{\alpha}(\overline{{\cal O}})$ when
$\overline{{\cal O}}={\cal O}_i$ due to 
the 
$I^{\ge} _{\alpha-1}({\cal O})$ term in the numerator of Eq. 
(\ref{eq40Adi}). This behavior is caused by long sticking times in 
a state of the system, on a time scale which is of the
order of the measurement time.

\section{Statistics of Weak Ergodicity Breaking} 
\label{SecStat}

In this Sec. we investigate properties of Eqs. (\ref{eq10},\ref{eq11}).

\subsection{Limits $\alpha \to 0$ and $\alpha \to 1$} 

In the limit
$\alpha \to 1$ we recover usual ergodic behavior. From Eq.  
(\ref{eq11})
\begin{equation}
f_{1} \left( \overline{{\cal O}} \right) =
- { 1 \over \pi} \lim_{\epsilon\to 0} \mbox{Im}
{ \sum_{i=1} ^L p^{{\rm eq}}_i \over 
 \sum_{i=1} ^L p^{{\rm eq}} _i \left( \overline{{\cal O}} - {\cal O}_i + i \epsilon\right)} .
\label{eq12}
\end{equation}
Using the normalization condition $\sum_{i=1}^L p^{{\rm eq}}_i=1$ and
the ensemble average 
\begin{equation}
\langle {\cal O} \rangle = \sum_{i=1} ^L p^{{\rm eq}}_i {\cal O}_i
\label{eq13}
\end{equation}
we have ergodic behavior
\begin{equation}
f_{1} \left( \overline{{\cal O}} \right) = \delta(\overline{{\cal O}} -\langle {\cal O} \rangle)
\label{eq14}
\end{equation}
in the sense that the ensemble average is equal to the time average. 
Note that already Eq. 
(\ref{eq02})
in the limit $\alpha \to 1$ indicates ergodicity since
the residence time is not a fluctuating quantity (i.e. the one sided
L\'evy PDF is a delta function when $\alpha =1$). 
In the opposite limit of $\alpha \to 0$ we find  
\begin{equation}
\lim_{\alpha \to 0} f_{\alpha} \left( \overline{{\cal O}} \right) =
- { 1 \over \pi} \lim_{\epsilon\to 0} \mbox{Im}
 \sum_{i=1} ^L p^{{\rm eq}}_i  
\left( \overline{{\cal O}} - {\cal O}_i + i \epsilon\right)^{-1}
\label{eq15}
\end{equation}
hence
\begin{equation}
\lim_{\alpha \to 0} f_{\alpha} \left( \overline{{\cal O}} \right) =
\sum_{i=1} ^L p_i ^{{\rm eq}} \delta \left(\overline{{\cal O}} - {\cal O}_i \right).
\label{eq16}
\end{equation}
This describes a localization behavior where the system is stuck in
one of the states for the whole duration of the observation time,
which is the
expected behavior when $\alpha \to 0$.

\subsection{Lamperti Statistics of the Occupation fraction} 

Let the Physical observable be
${\cal O}_i=1$ when $i=1,..., \bar{l}$ where 
$\bar{l}\le L$, otherwise ${\cal O}_i=0$.
Hence the time average in this case is 
the occupation fraction
\begin{equation}
\overline{{\cal O}}={\sum_{i=1}^{\bar{l}} t_i \over \sum_{i=1} ^L t_i} 
\label{eq17}
\end{equation}
which  is the fraction of time spent by the system in the observation
domain $i=1,\cdots, \bar{l}$. Clearly $0\le \overline{{\cal O}} \le 1$
in this
case.
 Using Eq. (\ref{eq11}) a straight 
forward calculation
gives
\begin{equation}
f_{\alpha} \left( \overline{{\cal O}} \right) = {1 \over \pi} { {\cal R} \left[ \left( 1 - \overline{{\cal O}} \right) \overline{{\cal O}} \right]^{\alpha -1} \sin \pi \alpha \over
{\cal R}^2 \left( 1 - \overline{{\cal O}} \right)^{2 \alpha} + \overline{{\cal O}}^{2 \alpha} + 2 {\cal R} \left[
\left( 1 - \overline{{\cal O}} \right) \overline{{\cal O}} \right]^\alpha \cos \pi \alpha }.
\label{eq18}
\end{equation}
This is the Lamperti PDF \cite{Lamp}
which is a natural generalization of
the arcsine distribution [the case $\alpha=1/2, {\cal R}=1$ 
Eq. (\ref{eqARC})].
The PDF Eq. (\ref{eq18})
has found 
several applications in non-ergodic systems mentioned in the introduction 
\cite{Marg,Bel1,Thaler,Euro,Japan,Lomholt}
and recently for the non-self averaging properties
of the quenched trap model \cite{Stas}.  
The parameter ${\cal R}$ in Eq. (\ref{eq18})
is called the asymmetry parameter and is given by
\begin{equation}
{\cal R} = {p^{{\rm eq}}_{1,\bar{l}} \over 1 - 
 p^{{\rm eq}}_{1,\bar{l}} } 
\label{eq19}
\end{equation}
where $p^{{\rm eq}}_{1,\bar{l}}=\sum_{i=1} ^{\bar{l}} p^{{\rm eq}} _i$
is the probability
in ensemble sense to be in the observation domain.
Since the observable attains two values ${\cal O}_i=1$ or   
${\cal O}_i=0$  the PDF Eq. 
(\ref{eq18}) has two divergences on $\overline{{\cal O}}=1$ and
$\overline{{\cal O}}=0$ which is a special case of the
more general rule discussed after Eq. 
(\ref{eq38ADIa}). 

\subsection{Low order moments of time averaged observables}

From the moment generating function $\hat{f}_{\alpha}(\xi)$ 
we can obtain moments of the time averages $\overline{{\cal O}}$
\begin{equation}
\langle \overline{{\cal O}}^n \rangle = \left( - 1 \right)^n {1 \over n!} {\partial^n  \over \partial \xi^n } \hat{f}_\alpha\left(\xi \right)|_{\xi =0 }.
\label{eq20}
\end{equation}
Using Eq. (\ref{eq10}) we find 
\begin{equation}
\langle \overline{{\cal O}} \rangle = \sum_{i=1}^L p^{{\rm eq}}_i {\cal O}_i. 
\label{eq21}
\end{equation}
The average $\langle .... \rangle$ is over an ensemble of realizations.
If the ensemble reaches an equilibrium then obviously 
$\langle \overline{{\cal O}} \rangle =  \langle {\cal O}\rangle$
which is time independent.
Hence  the 
$p^{{\rm eq}}_i$s in Eq. (\ref{eq21}) are the probabilities that
 a member of
an ensemble of systems occupies state $i$ when the ensemble reaches
an equilibrium. This justifies our original
assumption that the $p^{{\rm eq}}_i$s in Eq. 
(\ref{eq02}) are population fractions. 
Namely the $p^{{\rm eq}}_i$s can be in principle measured by letting many
independent systems (or many non-interacting particles) evolve and
then in the long time limit 
$p^{{\rm eq}}_i$ is the number of systems in state $i$ over
the total number of systems when the latter is large. 
Using Eqs.  (\ref{eq10},\ref{eq20}) the fluctuations
are given by
\begin{equation}
\langle \overline{{\cal O}}^2 \rangle - \langle \overline{{\cal O}}\rangle^2=
\left( 1 - \alpha \right) \left( \langle {\cal O}^2 \rangle - \langle {\cal O} \rangle^2 \right) 
\label{eq22}
\end{equation}
where $\langle {\cal O}^2 \rangle = \sum_{i=1} ^L p^{{\rm eq}}_i ({\cal O}_{i})^2$. Eq. (\ref{eq22}) gives a simple relation between fluctuations
of time averages and fluctuations of ensemble averages. 
Once again
when $\alpha = 1$ the fluctuations of the time average Eq. (\ref{eq22}) 
vanish, indicating ergodic behavior. 
 Relations between cumulants of time average observables 
and cumulants of ensemble averages are found in a similar way. 
For the third cumulant
\begin{widetext}

\begin{equation}
 C_{3} (\bar{\cal O} ) = \langle\bar{\cal O}^3\rangle -3 \langle\bar{\cal
O}^2\rangle \langle\bar{\cal O}\rangle +2 \langle\bar{\cal O}\rangle^3
= {1 \over 2} (2-\alpha)(1-\alpha)\left(\langle{\cal O}^3\rangle -3 \langle{\cal
O}^2\rangle \langle{\cal O}\rangle +2 \langle{\cal O}\rangle^3\right)
={1 \over 2} (2-\alpha)(1-\alpha)C_{3} ({\cal O} )
\label{EqCum3}
\end{equation}
and the fourth cumulant 
$$ C_{4} (\bar{\cal O} ) = \langle\bar{\cal O}^4\rangle -4 \langle\bar{\cal
O}^3\rangle \langle\bar{\cal O}\rangle-3\langle\bar{\cal O}^2\rangle^2
+12\langle\bar{\cal O}^2\rangle\langle\bar{\cal O}\rangle^2 -6 \langle\bar{\cal O}\rangle^4 =$$
\begin{equation}
 \left(1-\alpha\right)\left( \frac{(3-\alpha)(2-\alpha)}{6}\left(\langle{\cal
O}^4\rangle-4\langle{\cal O}^3\rangle\langle{\cal O}\rangle\right)
+ \left(6-6\alpha +\alpha^2\right)\left(\langle{\cal O}\rangle^2\left(2\langle{\cal
O}^2\rangle-\langle{\cal O}\rangle^2\right)\right) -
\frac{1}{2}(6-\alpha)(1-\alpha)\langle{\cal O}^2\rangle^2 \right).
\label{EqCum4}
\end{equation}
\end{widetext}
Low order moments of time averaged observables can be expressed   
using the cumulants Eqs. (\ref{eq22},\ref{EqCum3},\ref{EqCum4})
the results remain as cumbersome as the expressions in
Eqs. (\ref{EqCum3},\ref{EqCum4}).
When the odd moments of the ensemble average
are equal zero $\langle {\cal O}^{2 n -1} \rangle=0$ $n=1,2, \cdots$,
corresponding to examples we investigate in Sec. 
\ref{SecX},
the expressions
for moments are
simpler. Odd moments of the time average observable are  equal
zero, as expected. The second moment is 
$\langle \overline{{\cal O}}^2 \rangle=
(1 - \alpha) \langle {\cal O}^2  \rangle$ 
and the fourth moment 
\begin{equation}
 \langle \overline{{\cal O}}^4 \rangle = \frac{(3-\alpha)(2-\alpha)(1-\alpha)}{6} \langle
{\cal O}^4 \rangle
+ \frac{\alpha(1-\alpha)^2}{2}\langle {\cal O}^2 \rangle^2. 
\label{Eqfourth}
\end{equation}

\subsection{Correlations Between Occupation Fractions 
$\langle \overline{p}_l \overline{p}_k \rangle$} 

The correlations  between the occupation fractions
$\overline{p}_l = t_l / t $ and $\overline{p}_k=t_k/t$
where $t=\sum_{i=1} ^L t_i$  are now 
briefly investigated. We use the $L$ dimensional vector $\vec{\xi}=
\left\{\xi_1,\xi_2,\cdots \xi_L \right\}$ and 
the  $L$ dimensional generating function
\begin{equation}
\hat{g}_\alpha \left( \vec{\xi} \right) = \langle {1 \over 1 + \sum_{i=1} ^L \xi_i \overline{p}_i } \rangle.
\label{eq23}
\end{equation}
Related multidimensional arcsine distribution of occupation fractions
were investigated in \cite{Barlow}.
Using the substitution $\xi {\cal O}_i \to \xi_i$  in Eq. 
(\ref{eq03})
it is easy to see using Eq. (\ref{eq10})
\begin{equation}
\hat{g}_{\alpha}\left(\vec{\xi}\right) = { \sum_{i=1}^L p^{{\rm eq}}_i \left( 1 + \xi_i\right)^{\alpha - 1} \over \sum_{i=1}^L p^{{\rm eq}}_i \left( 1 + \xi_i \right)^\alpha}.  
\label{eq24}
\end{equation}
This multidimensional generating function yields
\begin{equation}
\langle \overline{p}_i \rangle = - {\partial \over \partial \xi_i} \hat{g}_\alpha \left( \vec{\xi} \right)|_{\vec{\xi} = 0 } = p^{{\rm eq}}_i
\label{eq25}
\end{equation}
and  similarly by taking the second order derivative of 
$\hat{g}_{\alpha}(\vec{ \xi})$
 with respect to 
$\xi_i$ 
\begin{equation}
\langle \overline{p}_i^2 \rangle - \langle \overline{p}_i\rangle^2 = 
\left( 1 - \alpha\right) p^{{\rm eq}}_i \left( 1 - p^{{\rm eq}}_i \right).
\label{eq26}
\end{equation}
Identical results can be obtained using the Lamperti PDF 
Eq. (\ref{eq18}).
More
interesting is to notice the correlations between occupation fractions
for example
\begin{equation}
\langle \overline{p}_l \overline{p}_k \rangle = {1 \over 2} {\partial \over \partial \xi_k } {\partial \over \partial \xi_l} \hat{g}_\alpha \left( \vec{ \xi } \right)|_{\vec{ \xi }=0}
\label{eq27}
\end{equation}
for $l \neq k$.
Using Eq. 
(\ref{eq24})
\begin{equation}
\langle \overline{p}_l \overline{p}_k \rangle = 
\alpha p^{{\rm eq}}_l p^{{\rm eq}}_k .
\label{eq28}
\end{equation}
We see that when $\alpha\to 1$ the occupation fractions
are uncorrelated  since
$\langle \overline{p}_l \overline{p}_k \rangle -
\langle \overline{p}_l\rangle \langle  \overline{p}_k \rangle = 0$
and one can show that they are independent random variables.
When $\alpha \to 0$
 the system occupies one state for practically
 the whole
duration of the measurement hence either $\overline{p}_l \simeq 1$ and
then obviously
 $\overline{p}_k \simeq 0$ or the opposite situation is found, or
both occupation fractions are zero (if $L>2$). In any case clearly the
product  $\overline{p}_l \overline{p}_k $ is zero when $\alpha \to 0$
and $l \neq k$  as
we have indeed found in Eq. (\ref{eq28}).

\section{Distribution of $\overline{x}$}
\label{SecX}

 We now consider  
a particle undergoing stochastic  fractional dynamics in a binding field.  
 The fractional Fokker--Planck equation 
\cite{MBK,PRE2}
describes anomalous sub-diffusion and
relaxation close to thermal equilibrium using fractional calculus
\begin{equation}
{\partial^\alpha p(x,t) \over \partial t^\alpha} = D_\alpha \left[ {\partial^2 \over  \partial x^2} - {\partial \over \partial x} {F(x) \over T} \right] p(x,t)
\label{eq31}
\end{equation}
where $D_\alpha$ is the fractional diffusion coefficient, and
$F(x) = - \partial V(x)/ \partial x$ is the force.  
Eq. (\ref{eq31})  reduces to the usual Fokker-Planck equation when $\alpha=1$. 
The fractional Fokker-Planck  equation was derived from the sub-diffusive
continuous time random walk \cite{PRE2}
 which is the stochastic process we have in mind.
In the absence of the
force field and for free boundary conditions
 $\langle x^2 \rangle = 2 D_\alpha t^{\alpha}$.
An important property of the fractional Fokker-Planck
equation is that in the long time
limit Boltzmann equilibrium is obtained \cite{MBK}
\begin{equation}
p^{{\rm eq}} \left( x \right) = 
{\exp\left[ - { V(x) \over T } \right] \over Z} 
\label{eq29}
\end{equation}
provided that the force is binding.  
Recently 
numerical methods which give
the sample paths of the fractional Fokker-Planck equation
were investigated in detail \cite{Goy,Mainardi,Friedrich,Friedrich1,Weron}.
Such paths or the corresponding CTRW trajectories yield
non ergodic behaviors \cite{Bel1}.
For example in \cite{BarkaiJSP} the Lamperti
PDF Eq. (\ref{eq18}) of the occupation fraction was obtained from the
fractional equation  
(\ref{eq31}). However so far distributions of time averages
of physical observables were not considered in detail. 

 We investigate the time average of the observable ${\cal O}=x$
with $-\infty < x< \infty$
so we are dealing with a continuum situation.
Taking the continuum limit of  Eq. 
(\ref{eq11})
we find
\begin{equation}
f_\alpha \left( \overline{x} \right) = - {1 \over \pi} \lim_{\epsilon \to 0} \mbox{Im} { \int_{-\infty} ^\infty {\rm d} x p^{{\rm eq}} \left(x\right) \left( \overline{x} - x  + i \epsilon\right)^{\alpha - 1} \over \int_{-\infty} ^\infty {\rm d} x p^{{\rm eq}} (x) \left( \overline{x} - x + i \epsilon\right) ^\alpha } 
\label{eq32}
\end{equation}
which for $0<\alpha<1$ is rewritten as
%
%\begin{widetext}
%\begin{equation}
%f_\alpha \left( \overline{x} \right) = - {1 \over \pi} \mbox{Im} { \int_{-\infty} ^{\overline{x}} {\rm d} x p^{{\rm eq}} (x) \left( \overline{x} - x  \right)^{\alpha - 1} + \int_{\overline{x}} ^{\infty} {\rm d} x p^{{\rm eq}} (x) \left(x- \overline{x} \right)^{\alpha - 1} e^{i \pi \left( \alpha -1\right)}  \over 
%\int_{-\infty} ^{\overline{x}}  p^{{\rm eq}} (x) \left( \overline{x} - x \right) ^ \alpha + \int_{\overline{x}} ^{\infty} p^{{\rm eq}} \left( x \right) \left(x-  \overline{x} \right)^\alpha e^{ i \pi \alpha} } 
%\label{eq33}
%\end{equation}
%
%And finally
%
 $$ f_{\alpha} \left( \overline{x} \right) =  $$
\begin{equation}
{\sin \pi \alpha \over \pi}  
{ I_{\alpha -1} ^< \left( \overline{x} \right) I_{\alpha} ^> \left( \overline{x} \right) + I^{>} _{\alpha - 1} \left( \overline{x} \right) I_{\alpha} ^{<} \left( \overline{x} \right) 
\over
\left[ I_{\alpha} ^{>} \left( \overline{x} \right) \right]^2 + \left[ I^{<} _\alpha \left( \overline{x} \right) \right]^2 + 2 \cos \pi \alpha I^{>} _\alpha \left( \overline{x} \right) I^{<} _\alpha \left( \overline{x} \right) },
\label{eq34}
\end{equation}
%\end{widetext}
%
where
\begin{equation}
I^{<} _{\alpha} \left( \overline{x} \right) = \int_{\overline{x}} ^\infty {\rm d} x p^{{\rm eq}} (x) |x- \overline{x}|^\alpha
\label{eq35}
\end{equation}
and
\begin{equation}
I^{>} _{\alpha} \left( \overline{x} \right) = \int_{-\infty} ^{\overline{x}} {\rm d} x p^{{\rm eq}} (x) |\overline{x} - x|^\alpha
\label{eq36}
\end{equation}
and similarly for $I^{<}_{\alpha -1} \left( \overline{x} \right)$ and
$I^{>}_{\alpha -1} \left( \overline{x} \right)$.
When $\alpha \to 0$ we have 
$\lim_{\alpha\to 0} f_{\alpha}\left( \overline{x} \right) = p^{{\rm eq}} (\overline{x}) $
 which is the continuum limit of Eq. 
(\ref{eq16}). In the ergodic limit $\alpha \to 1$
we find 
$f_1\left( \overline{x} \right)= 
\delta\left( \overline{x} - \langle x \rangle \right)$. 

\subsection{Free particle.} 

As an example consider a particle in a domain $-l/2<x<l/2$ undergoing
an unbiased fractional random walk with reflecting walls.
 This is a free particle in the sense that
no external field is acting on it. The time average of the particle's
position $\overline{x}$ is considered, and obviously for this
case $p^{{\rm eq}}(x) = 1/l$ for $-l/2<x<l/2$. Using Eq.
(\ref{eq34}) we find the PDF of  $\overline{x}$
$$ f_{\alpha} \left( \overline{x} \right) = $$
\begin{equation}
{ 1 \over l}
{ N_\alpha  \left( { 1 \over 4} - { \overline{x}^2 \over l^2} \right)^\alpha  \over
\left| { 1 \over 2} - { \overline{x} \over l} \right|^{ 2 ( 1 + \alpha) } +
\left| { 1 \over 2} + { \overline{x} \over l} \right|^{ 2 ( 1 + \alpha) } +
2 \left| { 1 \over 4} -  \left( { \overline{x} \over l} \right)^2 \right|^{  1 + \alpha } \cos \pi \alpha },
\label{eq37}
\end{equation}
where $N_\alpha=( 1 + \alpha ) \sin \pi \alpha / (\pi \alpha )$.
When $\alpha \to 1$ we have  ergodic behavior
$f_{\alpha}\left(\overline{x}\right) =\delta(\overline{x})$ since
 $\langle x \rangle=0$ while
in the opposite limit 
$f_{\alpha \to 0} \left( \overline{x} \right) = 1/l $ for $|\overline{x}|<l/2$
which is the uniform distribution, reflecting the mentioned
localization of the
particle in space when $\alpha \to 0$.

\begin{figure}
\begin{center}
\epsfxsize=80mm
\epsfbox{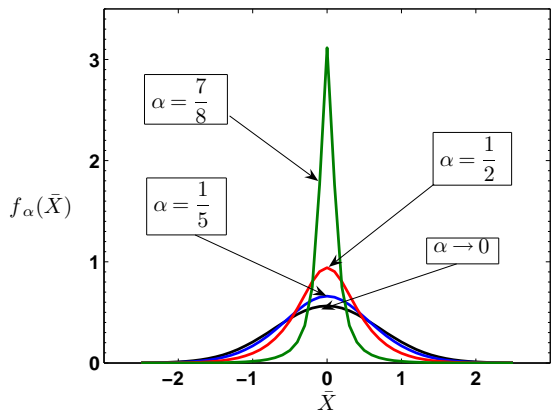}
\end{center}
\caption{
The PDF of $\overline{x}$ for a particle in a Harmonic force field
Eq. 
(\protect\ref{eq34}). 
We find a transition between an ergodic behavior:
 a delta distribution of $\overline{x}$ when
$\alpha \to 1$,  to the localization limit where the distribution
of $\overline{x}$  is Gaussian when $\alpha \to 0$.  
}
\label{fig1}
\end{figure}

\subsection{Harmonic Oscillator} 

We consider the time average
$\overline{x}$ of a particle
in a Harmonic force field 
$V(x)/T = x^2$ so that $p^{{\rm eq}}(x) = \exp( - x^2)/\sqrt{\pi}$ 
and $\langle x \rangle=0$.
Using Mathematica the integrals 
Eqs. (\ref{eq35},\ref{eq36}) 
can be calculated explicitly and expressed
in terms of tabulated confluent Hypergeometric functions. 
 In Fig. \ref{fig1}
the PDF of $\overline{x}$ Eq. (\ref{eq34}) is presented,  
and  a transition from a narrow 
distribution when $\alpha \to 1$ to 
a Gaussian distribution 
when $\alpha \to 0$,
$\lim_{\alpha \to 0} f_{\alpha} \left( \overline{x} \right) = 
p^{{\rm eq}}\left(\overline{x} \right)$ is found. 
Using Eq. (\ref{eq34}) it is easy to show
that 
$\langle \overline{x} \rangle=0$, $\langle \overline{x}^2 \rangle= 
(1 - \alpha)\langle x^2 \rangle$ with $\langle x^2 \rangle= 1/2$ for 
$p^{\rm eq }(x)$ under consideration, and $\langle \overline{x}^4 \rangle =
(1 - \alpha) (3 - 2 \alpha) \langle x^2 \rangle^2$. Only when $\alpha \to 0$
we have Gaussian statistics  with $\langle \overline{x}^4 \rangle =
 3 \langle \overline{x}^2 \rangle^2$. 
The PDF $f_{\alpha}(\overline{x})$ at its maximum on
 $\bar{x}=0$ is 
\begin{equation}
 f_{\alpha}\left(\overline{x}=0\right)=\frac{ \Gamma(\frac{\alpha}{2}) \tan(\frac{\pi\alpha}{2})
}{\Gamma(\frac{1+\alpha}{2}) \pi  },
%=\frac{ 1}{
%\cos(\frac{\pi\alpha}{2})\Gamma(1-\frac{\alpha}{2}) \Gamma(\frac{1+\alpha}{2})
%}
\label{eqATzer}
\end{equation}
which is equal to $p^{{\rm eq}}(x=0)=1/\sqrt{\pi}$ when $\alpha \to 0$ and
diverges when $\alpha \to 1$ as expected from an ergodic behavior.
For the Harmonic oscillator and the Free particles
the maximum of $f_{\alpha}( \overline{x})$
is found on the ensemble average $\langle x \rangle =0$, so the most likely
result for $\overline{x}$ is $\langle x \rangle$. In the
next subsection we consider a case where 
a minimum of $f_{\alpha}( \overline{x})$ is found on the ensemble average. 
                                                                                
\subsection{Double well potential.} 

\begin{figure}
\begin{center}
\epsfxsize=80mm
\epsfbox{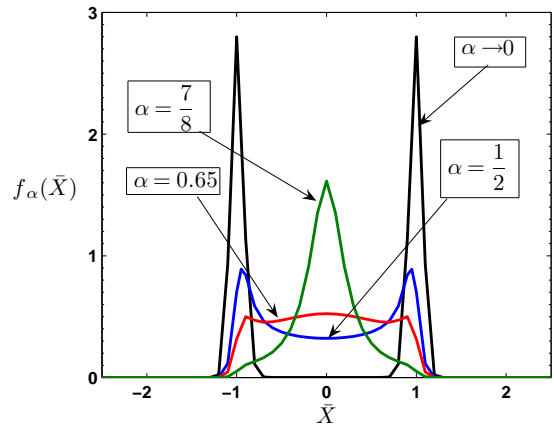}
\end{center}
\caption{
The PDF of $\overline{x}$ for a particle in the double well potential 
with $T=0.01$. A transition between a bimodal behavior 
when $\alpha< \alpha_c$ to a PDF with a peak on $\overline{x}=0$ when 
$\alpha>\alpha_c$ is observed ($\alpha_c \simeq 0.59$ for this case). 
In the ergodic limit $\alpha \to 1$
$f_{\alpha} \left( \overline{x} \right)$ is a delta function 
centered on the ensemble average 
$\langle x \rangle =0$.
 In the localization limit $\alpha \to 0$
 $f_{\alpha} \left( \overline {x} \right)$ is equal
to the population density $p^{{\rm eq}} (\overline{x})$. 
}
\label{figay}
\end{figure}

\begin{figure}
\begin{center}
\epsfxsize=80mm
\epsfbox{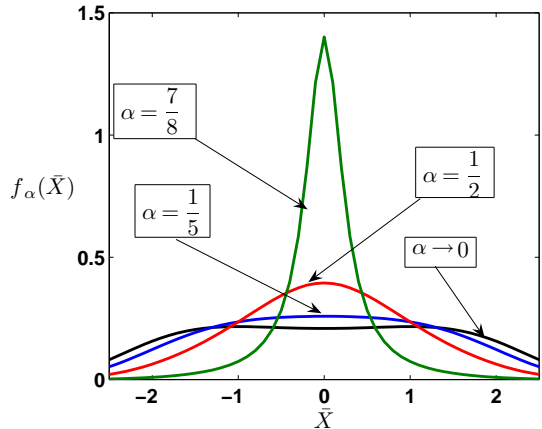}
\end{center}
\caption{
The same as Fig. (\ref{figay}) however now $T=7$. 
The bimodal shape presented in Fig. (\ref{figay}) is smoothed and we
barely observe bi-modal behavior, since
the equilibrium density $p^{{\rm eq}}(x)$ is not
centered around the two minima of the double well potential,
when the temperature is high. 
}
\label{figaz}
\end{figure}

An interesting case is the symmetric double well potential 
$V(x)/T = (x^4/4 - x^2/2)/T$ so $\langle x \rangle=0$.
  When $T \to 0$, $p^{{\rm eq}}(x)$
has two peaks  centered on the two local minima
of the double well potential. 
In this low 
temperature case
and in the limit $\alpha \to 0$ we expect to find the particle either
in the left well or in the right well for a time scale
 comparable to the measurement time. Hence when
$T \to 0$ and $\alpha \to 0$ the
PDF of the time average $\overline{x}$, $f_\alpha\left( \overline{x} \right)$
is a sum of two delta functions
since either $\overline{x}=1$ or 
$\overline{x}=-1$.
When $\alpha \to 1$ 
we expect an ergodic behavior, and then
PDF $f_1\left( \overline{x} \right) = \delta\left( \overline{x} \right)$,
since $\langle x \rangle=0$. 
So for low temperatures we expect a transition in
the behavior of $f_\alpha \left( \overline{x} \right)$ from a bimodal shape
when $\alpha \to 0$ to a PDF with a single peak 
centered on  zero when $\alpha \to 1$. 
Hence we will have a critical value $\alpha_c$. 
For $\alpha < \alpha_c$ the shape of $f_\alpha\left( \overline{x} \right) $
is bimodal with a minimum on $\overline{x}=0$,
while for $\alpha>\alpha_c$ a maximum on $\overline{x}=0$ is found.
These low temperature behaviors are shown in Fig. 
\ref{figay}.
For high temperatures (compared with the barrier height)
the bimodal solution of $p^{{\rm eq}}(x)$ turns into a
flatter shape.  
Since
$\lim_{\alpha \to 0} f_{\alpha} \left( \overline{x} \right) = p^{{\rm eq}}(\overline{x})$
we will not observe the bimodal shape of 
$\lim_{\alpha \to 0} f_{\alpha} \left( \overline{x} \right)$
when $T \to \infty$. 
Such high temperature behavior is shown in Fig. \ref{figaz}. 

\begin{figure}
\begin{center}
\epsfxsize=80mm
\epsfbox{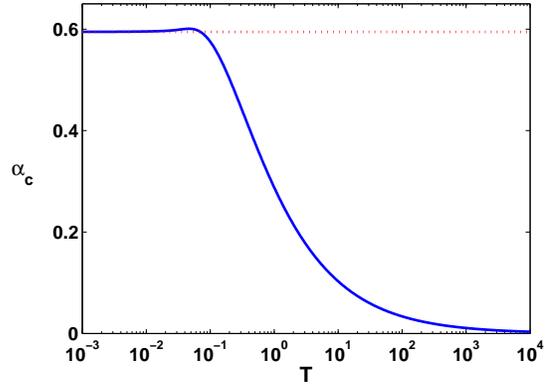}
\end{center}
\caption{
The critical exponent $\alpha_c$ versus temperature for a particle in
a double well potential. 
$\alpha_c$ marks the
transition between a local minimum to a local maximum of the PDF of
 $\overline{x}$ 
on $\overline{x}=0$.  When $T \to \infty$ $\alpha_c \to 0$ since 
in this limit and when $\alpha \to 0$ $f_{\alpha} \left( \overline{x} \right)$
is equal to the population density
$p^{{\rm eq}} \left( \overline{x}\right)$ which does not
``feel" the wells when the temperature is high (single peak).
When $T \to 0$ we have $\alpha_c \simeq 0.595$ . 
}
\label{figaa}
\end{figure}

  Investigating the extremum of $p^{{\rm eq}}(x)$ on $x=0$ 
it is easy to show that $\alpha_c$ is finite for any finite
temperature 
 and $\alpha_c \to 0$ when $T \to \infty$. 
For $T=0$ we have only two states in the system, either
$x=-1$ or $x=1$ corresponding to two minima of the double
well potential. The analysis is then very similar to the
two state ballistic L\'evy walk model \cite{ZK,Marg}.
 Clearly   $\overline{x}$ is the residence time in
state $x=1$ minus the residence time in state $x=-1$ divided by the
measurement time $t$.  Since
the sum of these two residence times is the measurement
time we can use the Lamperti  distribution Eq. (\ref{eq18})
to predict the distribution of $\overline{x}$. So when $T \to 0$
$$ \lim_{T \to 0} f_{\alpha} \left( \overline{x} \right) = $$
\begin{equation}
{ 2 \sin \pi \alpha \over \pi} { \left( 1 - \overline{x}^2 \right)^{\alpha -1}  \over
\left( 1 + \overline{x}\right)^{2 \alpha} +
 \left( 1 - \overline{x}\right)^{2 \alpha} +
2 \left( 1 - \overline{x}^2 \right)^\alpha \cos \pi \alpha} 
\label{eqMAG}
\end{equation}
which was found already in \cite{Godreche,Bald}. 
We find  that $\alpha_c=0.59461\cdots$ when $T \to 0$.
The behavior of $\alpha_c$ versus temperature is shown
in Fig.  \ref{figaa} and the transition
between the low and high temperature cases is presented.

\subsection{Possible Physical Applications}

 It is interesting to verify in experiments our
theoretical predictions and here we discuss three
examples.   Generally systems with CTRW type of dynamics
are natural candidates for the investigation of weak ergodicity
breaking, provided that information of single particle dynamics
can be recorded. 

 {\em Sub-diffusion $\langle x^2 \rangle \sim t^\alpha$
of a bead in a polymer network}
was measured by Wong et al
\cite{Weitz}.
The measured \cite{Weitz} exponent $\alpha$  
depends on the ratio of the size of the bead
and the linear size of the mesh of the network $l$ (roughly a $\mu$m).
We suggest to add an external binding field,
for example an harmonic
trap.
The time averages  of  a single particle coordinate can then be 
measured,
and according to our theory  its distribution is
given by Eq. (\ref{eq11}).
Such measurement can provide  insight into the nature
of disorder, for example is it quenched or annealed \cite{Stas}.

{\em Messenger RNA molecules inside live E. coli cells}
 exhibit anomalous diffusion
$\langle x^2 \rangle \sim t^\alpha$ and $\alpha =3/4$ \cite{Cox}.
 Due to the finite size
of the cell the motion is bounded. It would be interesting to investigate
time averages of the position of the single molecule, or occupation
time statistics, to investigate deviations from ergodicity.
Our theory gives a prediction for the distribution of these
observables, which can be tested in experiment.

{\em Blinking quantum dots} exhibit ergodicity breaking which is 
already measured in experiments \cite{Dahan,Marg1,Marg3}.
So far a simple two state picture
of the quantum dots was used, either the dot is on and it emits
many photon, or it is off \cite{Marg}.
Then ergodicity breaking of the time averaged
fluorescence intensity is similar to the time average of a particle in the
double well potential in the limit
of $T \to 0$, in the latter case
either the particle is the left well or in the
right well. More careful analysis reveals that some dots
deviate from a simple two state process \cite{Marg3}.
Then our more general
theory can be used in principle to predict distribution of 
time averaged intensity 
beyond the existing two state approach.    

\section{From Continuous Time Random Walk to Weak Ergodicity Breaking}
\label{SecCTRW}

In this Sec.  we derive our main results using the CTRW approach.
To reach  
Eq. (\ref{eq11}) 
we assumed among other things  that: (i) the
PDF of the occupation time $t_i$ is the one sided  L\'evy PDF
Eq. (\ref{eq02}) 
and (ii) that the total measurement
time $t = \sum_{i=1}^L t_i$ is a random variable,
while in Physical experiments
the measurement time is fixed. 
These assumptions are relaxed now using two types of CTRW models. 
Thermal CTRW describe
a Physical situation where the particle is undergoing the random 
walk in contact with a thermal heat bath. In this case the
equilibrium distribution of an ensemble of particles is 
Boltzmann's distribution. The second case describes a
system far from thermal equilibrium, where  a non-thermal
equilibrium is reached. 

We consider a renewal process for a system with $L$ states $i=1,...,L$.
The system starts in state $i$, it waits in this state until
time $\bar{t}_1$, it then jumps to some other state say state $l$,
it waits in state $l$ until time $\bar{t}_2$ 
and then makes another
jump. 
The sojourn times between jump events $\tau$ are
independent identically distributed random variables with a common
PDF $\psi(\tau)$. Our focus is on the case where $\psi(\tau)$ has
a long tail $\psi(\tau) \propto \tau^{-(1 + \alpha)}$ when $\tau \to \infty$
so $\langle \tau \rangle = \infty$ when $0<\alpha<1$.
After waiting in a state $i$ a transition to state
$j \ne i$ takes place,  with  probability $w_{ji}$ 
($0\le w_{ji} \le 1$ , $\sum_{j=1} ^{L} w_{ji}=1$ and $w_{ii}=0$).  
We assume that transition probabilities $w_{ji}$ are 
such that in the limit of long measurement times
all states are visited what ever
is the initial condition. In other words the system is not
decomposable into non-accessible regions in the space it samples,
 where once the system
starts in a certain region it cannot explore all other states. Such
a case corresponds to strong ergodicity breaking.  

\begin{figure}
\begin{center}
\epsfxsize=80mm
\epsfbox{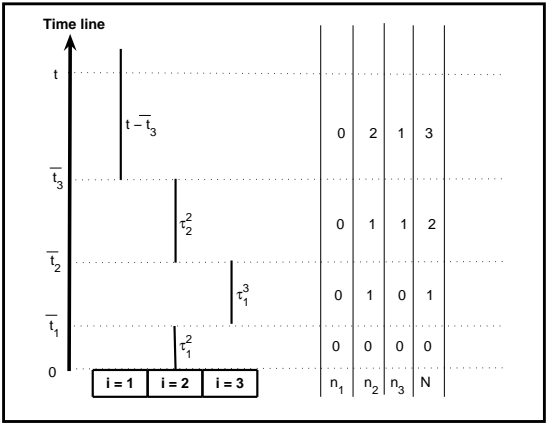}
\end{center}
\caption{
A schematic diagram of the process for a system
with three states, starting in state $2$ and
ending in state $k=1$. In this example the occupation times
are $t_1=t-\bar{t}_3$, $t_2 = (\bar{t}_1-0)+ (\bar{t}_3 - \bar{t}_2)=
\tau_{1} ^2 + \tau_2 ^{2}$ and $t_3= \bar{t}_2 - \bar{t}_1= \tau_{1} ^{3}$.
}
\label{fig2}
\end{figure}

 Let $N$ be the random number of jump events (renewals) in the
time interval $(0,t)$. Dots on the time axis on which jumps from
one state to another happen are denoted with $\bar{t}_i$ and clearly
$\bar{t}_N < t < \bar{t}_{N+1}$. Let $n_i$ be the number of transitions
out of state $i$ and clearly $N=\sum_{i=1} ^L n_i$. 
Let $\tau_{l} ^{i}$ be the $l$ th sojourn time in state $i$.
And let $k$ be the state of the system at time $t$. A
schematic presentation of the process with three states is
shown in Fig. \ref{fig2}.
Statistics of the number of renewals $N$ in $(0,t)$ is a well investigated
problem \cite{Feller,Godreche,Shlesinger} for example
$\langle N \rangle \sim t^\alpha$.

For the CTRW on a one dimensional lattice the states $i=1,...L$  
correspond to the position of the particle on a finite lattice.
 Then for
example 
the time average of the coordinate of the particle is 
$\overline{x}=\sum_{i=1} ^L i t_i / t$ where $t_i$ is the occupation 
time in state $i$ (see Fig. 
\ref{fig2}).
 However our considerations are more general.
For example for blinking quantum dots
\cite{Dahan,Marg,Marg1,Marg3}
one state (say $i=1$) may denote an on
state in which many photons are emitted and state $i=2$ is the off state.
This system is non-thermal since it is driven
by a strong laser field. On the other hand the CTRW dynamics of
a probe bead immersed in a polymer actin network \cite{Weitz}
is an example
for a thermal CTRW motion in a system with
a well defined temperature $T$.

 A specific example is the CTRW on a
lattice with jumps to nearest neighbors
only.
A particle on $i$
has a probability of jumping left $q_i$ and a probability of jumping
right $1-q_i$ so
\begin{equation}
 w_{i-1,i}=q_i \ \ \ \mbox{and} \ \ \ \  w_{i+1,i} = 1 - q_i .
\label{eqtran}
\end{equation}
Reflecting boundary conditions $q_L=1$ and $q_1=0$ are assumed.
For $i\ne 1,L$ we must have $q_i \neq 0$ and $q_i \neq 1$ 
so that all lattice points be visited. 
Between jumps the particle waits on a lattice point.
The waiting times between the jumps are 
independent, identically distributed random variables with
a common PDF $\psi(\tau)$. 
This type of random walk leads to anomalous subdiffusion
$\langle x^2 \rangle \sim t^\alpha$ when $q_i=1/2$ and the system
is infinite \cite{SM,Shlesinger}. 
%
%\begin{equation}
%p_i \left(N+1\right) = q_{i+1}p_{i+1} \left( N \right) + \left( 1 -  q_{i-1} \right) p_{i-1} \left( N \right).
%\label{eqDIS}
%\end{equation}
%

 The ratio  $v_i=n_i/N$ is called the visitation fraction. 
The population fraction $p^{{\rm eq}}_i$ is found by considering 
the ensemble of $M$ non interacting systems. Letting these systems
evolve from some initial condition and waiting for the long
time limit,  $\lim_{M \to \infty} M_i /M =  p^{{\rm eq}}_i$
where $M_i$ is the number of systems in state $i$. 
The population fraction 
is determined from $w \cdot  p^{{\rm eq}}= p^{{\rm eq}}$. 

After many jumps $N \to \infty$ 
and for any initial condition
the visitation fraction reaches an equilibrium
and  
\begin{equation}
p^{{\rm eq}}_i= \lim_{N \to \infty} v_i= v_i ^{{\rm eq}}
\label{eqstar}
\end{equation}
so  $w \cdot v=v$.  
To see this note that the visitation fraction is given
by $v_i = \sum_{n=1} ^N \theta_i (n)/N$, where $n$ is a counter
of the number of jumps,
 and $\theta_i (n) = 1$ if the
particle is on $i$ after $n$ steps, otherwise it is zero. 
 In the
long time limit the PDF of $v_i $ will converge to a narrow
distribution centered around its mean so
$v_i \simeq \sum_{n=1} ^N \langle \theta_i (n) \rangle/N$.
Let $p_i (n)$ be the probability to be on $i$ after $n$ steps. 
By definition $\theta_i (n) = 1$ with probability $p_i (n)$
and $\theta_i (n) =0$ with probability $1 - p_i (n)$.
Hence
\begin{equation}
 v_i \simeq {\sum_{n=1}^N  p_i (n) \over N}
\label{eqvi}
\end{equation}
or in vector  notation $v=(v_1,\cdots,v_L)$, $p(n)=(\cdots,p_i(n) \cdots)$
we have $v\simeq  \sum_{n=1}^N p(n)/N$. 
This means that ergodicity holds in discrete time, where the operational
time is the number of steps, not the real time.
Hence the term {\em weak} ergodicity breaking \cite{WEB}
is very appealing. 
Multiplying 
Eq. (\ref{eqvi})
with $w$ from the left 
and using  $p(n+1)= w p(n)$ 
we have  $w \cdot v \simeq \sum_{n=1}^N p(n+1) /N $.
Hence when
 $N \to \infty$
we have $ w \cdot v =v$ which holds in the long time limit. 
It is important to realize
that the visitation fraction and the population fraction are equal
since all sojourn times have a common distribution
$\psi(\tau)$. We will later consider the more general case
where different states may have different waiting times PDFs. 

 For the one dimensional CTRW on a lattice the equilibrium
population and hence the visitation fraction is determined from 
Eq. (\ref{eqtran})
\begin{equation}
p^{{\rm eq}} _{i} = q_{i+1}p^{{\rm eq}} _{i+1} + \left( 1 -  q_{i-1} \right) p^{{\rm eq}} _{i-1}.
\label{eqDISa}
\end{equation}
Using reflecting boundary conditions
and Eq. (\ref{eqDISa})  
\begin{equation}
 p^{{\rm eq}}_i = 
\lim_{N \to \infty} v_i =
{1 \over  1-q_i} \Pi_{k=2} ^i {1-q_k\over q_k} 
p^{{\rm eq}}_1  
%\lim_{N \to \infty}  v^{{\rm eq}}_1
\label{eqEquil}
\end{equation}
and from normalization
\begin{equation}
p^{{\rm eq}}_1 = 
v^{{\rm eq}}_1 = 
\left[ 1 + \sum_{i=2}^L {1 \over 1 - q_i} \Pi_{k=2} ^i {1 - q_i \over q_i} \right]^{-1}.
\label{eqEquila}
\end{equation}
When the particle undergoing the CTRW process is coupled to a thermal
heat bath, we apply usual detailed balance condition on the transition
probabilities \cite{Bel2}. 
In this case the visitation fraction will be described
by Boltzmann statistics. For example if $q_i$
 is a constant $q>1/2$ the random
walk is biased, which Physically corresponds to an external force field
$F<0$ driving the particles to the left.
 Using lattice spacing $a$ and letting the system
be semi-infinite $L \to \infty$, thermal detailed balance condition 
gives the
ratio between the probability of  jumping left from point
$i$ and the probability of  jumping right from point $i-1$
\begin{equation}
{q_i \over 1 - q_{i-1}}= {q \over 1 - q}= \exp({|F| a \over T}  ).
\label{eqDB}
\end{equation}
Using 
Eqs. (\ref{eqstar},\ref{eqEquil},\ref{eqEquila},\ref{eqDB})
Boltzmann's statistics holds both  for the visitation 
and the population fractions
\begin{equation}
\lim_{N \to \infty} v_i = p^{{\rm eq}}_i =   { \exp\left( - {  E_i \over T } \right) \over Z}
\label{eqDB1}
\end{equation}
for $i>1$ where $E_i = |F| a i$ is the potential energy,
$Z$ is a normalization and for
the reflecting boundary $p^{{\rm eq}}_1=[1 - \exp( - |F| a / T)]/2$. 
More general thermal detailed balance conditions \cite{Bel2}
show that Eq. (\ref{eqDB1})
is valid for binding force fields and not limited 
to the case $F$ being
a constant. 
 
 The time average of a physical observable is as before
\begin{equation}
\overline{{\cal O}} = {\sum_{i=1} ^L {\cal O}_i t_i \over t},
\label{eq40}
\end{equation}
where now the measurement time $t$ is fixed and $\sum_{i=1} ^L t_i = t$. 
Let $\vec{{\cal O}} = \left\{ {\cal O}_1 , \cdots {\cal O}_L \right\}$.
We consider the moment generating function, 
\begin{equation}
\hat{f}_{t,\vec{{\cal O}} } \left( u \right)  = \langle \exp\left( - u \sum_{i=1} ^L
{\cal O}_i t_i  \right) \rangle,
\label{eq41}
\end{equation}
and in  double Laplace space
\begin{equation}
\hat{f}_{s,\vec{{\cal O}} } \left( u \right)  = \int_0 ^\infty e^{ - s t}  \langle \exp\left( - u \sum_{i=1} ^L
{\cal O}_i t_i  \right)\rangle {\rm d} t
\label{eq42}
\end{equation}
so $s,t$ and $u, \sum_{i=1} ^L{\cal O}_i t_i $ are two Laplace pairs.
Let $\vec{n} = \left\{ n_1, \cdots , n_L \right\}$.
We consider the generating function conditioned that the system
made $N$ transitions and $\vec{n}$ describes 
the number of renewals in each state. 
The occupation time in state $i \neq k$
is
\begin{equation}
t_i = \sum_{l=1} ^{n_i} \tau_{l} ^{i}
\label{eq44}
\end{equation}
and for state $k$
\begin{equation}
t_k = t - \bar{t}_N + \sum_{l=1} ^{n_k} \tau_{l} ^{k}. 
\label{eq45}
\end{equation}
The time $t-\bar{t}_N$ is called the backward recurrence time,
it is the time between the last jump event in $(0,t)$ and
the measurement time $t$. 
Using Eqs. (\ref{eq44},\ref{eq45}) 
the conditioned generating function is 
\begin{widetext}
\begin{equation}
\hat{f}_{s,\vec{{\cal O}} , N,  \vec{n}} \left( u \right) =
\langle \int_0 ^\infty {\rm d} t \exp\left[ - s t - u {\cal O}_k \left( t - \bar{t}_N +
\sum_{l=1} ^{n_k} \tau_{l} ^k \right) - \sum_{i=1,i \neq k} ^{L} u {\cal O}_i \sum_{l=1} ^{n_i} \tau_{l} ^{i} \right] I\left( \bar{t}_N < t < \bar{t}_{N+1} \right) \rangle
\label{eq46}
\end{equation}
where $I(x) =1$ if the condition in the parenthesis is true,
other wise $I(x) = 0$. 
First we integrate over $t$ and obtain
\begin{equation}
\hat{f}_{s,\vec{{\cal O}} , N,  \vec{n}} \left( u \right) =
\langle { e^{ - s \bar{t}_N} - e^{ - s \bar{t}_{N+1} - u {\cal O}_k \left(
\bar{t}_{N+1} - \bar{t}_N \right) } \over
s + u {\cal O}_k } e^{ - \sum_{i=1} ^L u {\cal O}_i \sum_{l=1} ^{n_i} \tau_{l} ^{i} } \rangle,  
\label{eq47}
\end{equation}
then we use
the assumption  of independent and identically distributed sojourn times
$\tau$,
and the identities
 $\bar{t}_N = \sum_{i=1}^L \sum_{l=1} ^{n_i} \tau_{i} ^{l}$,
$\bar{t}_{N+1} = \bar{t}_N + \tau_k ^{n_k +1}$  
to find
\begin{equation}
\hat{f}_{s,\vec{{\cal O}} , N,  \vec{n}} \left( u \right) = {\Pi_{i=1}^L \hat{\psi}^{n_i} \left( s + u {\cal O}_i \right) \left[ 1 - \hat{\psi}\left( s + u {\cal O}_k \right) \right] \over s + u {\cal O}_k }, 
\label{eq48}
\end{equation}
\end{widetext}
where  
$\hat{\psi}(s)=\int_0 ^\infty \psi(\tau) \exp( - s \tau) {\rm d} \tau$
is the Laplace transform of $\psi(\tau)$.

 In the limit of long measurement time, corresponding to the usual
small $s$ and $u$ limit, their ratio being finite, the system will 
reach an equilibrium for the number  of renewals in each state.
Namely from Eq. (\ref{eqstar}) the visitation fraction will satisfy
\begin{equation}
v_i = \lim_{N \to \infty} {n_i \over N } = p^{{\rm eq}} _i .
\label{eq49}
\end{equation}
We use Eq. (\ref{eq49}) in Eq. (\ref{eq48}), then insert the usual
small 
$s$ behavior  \cite{Metzler,Feller,Godreche,Shlesinger}
\begin{equation}
\hat{\psi}(s) \sim 1 - A s^{\alpha}
\label{eq49a}
\end{equation}
which corresponds to
$\psi(\tau) \sim A \tau^{- (1 + \alpha)}/|\Gamma(-\alpha)|$
and find
\begin{equation}
\hat{f}_{s,\vec{{\cal O}},k } \left( u \right) \sim { \left( s + u {\cal O}_k \right)^{\alpha - 1} \over
\sum_{i=1} ^L p_{i} ^{{\rm eq}} \left( s + u {\cal O}_i \right)^\alpha } .
\label{eq50}
\end{equation}
Summing over all final states $k$, with the final weights,  we find
\begin{equation}
\hat{f}_{s, \vec{{\cal O}} } \left( u \right) \sim 
{ \sum _{i=1} ^L p^{{\rm eq}}_i 
\left( s + u {\cal O}_i \right)^{\alpha-1} \over 
\sum_{i=1} ^L p_i ^{{\rm eq}} \left( s + u {\cal O}_i \right)^\alpha }.
\label{eq51}
\end{equation}
Using a method found in the Appendix of 
Ref. \cite{Godreche}  we invert Eq. (\ref{eq51}) and
find our main Eq. 
(\ref{eq11}).

\begin{figure}
\begin{center}
\epsfxsize=80mm
%\epsfbox{articlefigure3.eps}
\epsfbox{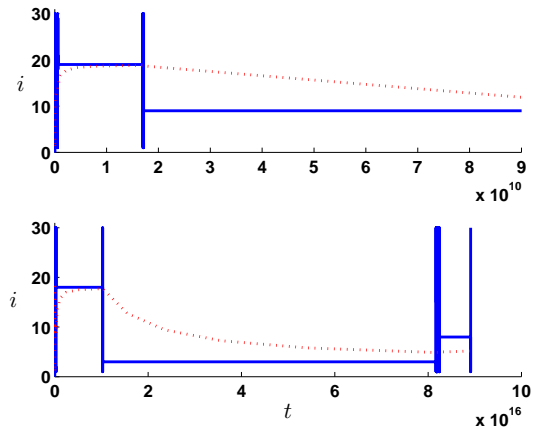}
\end{center}
\caption{
Trajectory of a single CTRW particle on a lattice, with $\alpha=0.3$
(solid curve, blue online). Long waiting times, of the order of the
measurement time, dominate the landscape. 
The time average $\overline{x}$ is a random
variable (dotted red curve). 
}
\label{figa03}
\end{figure}

\begin{figure}
\begin{center}
\epsfxsize=80mm
%\epsfbox{articlefigure3.eps}
\epsfbox{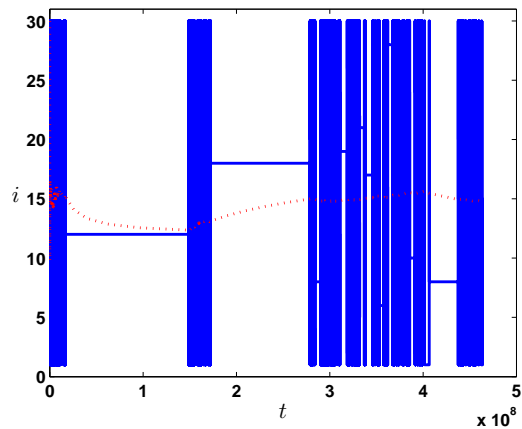}
\end{center}
\caption{
Same type of non-ergodic behavior
as found in Fig. \protect{\ref{figa03}} however now $\alpha=0.75$. 
}
\label{figa75}
\end{figure}

\subsection{Numerical Examples}
\label{subSnum}

 To demonstrate our results we consider an unbiased CTRW. We consider
a model
with $L=30$ sites on $i=1,...30$ jumps are to nearest neighbors only 
with $q_i = 1/2$
with periodic boundary conditions. 
We used the waiting time PDF
$\psi(\tau) = \alpha \tau^{- \alpha -1}$ for $\tau > 1$ otherwize
$\psi(\tau) = 0$. 
Simulating trajectories
of a single particle
we calculate the time
average $\overline{x}$ 
and then repeat the experiment many times and construct the distribution
of $\overline{x}$. 

\begin{figure}
\begin{center}
\epsfxsize=80mm
%\epsfbox{articlefigure3.eps}
\epsfbox{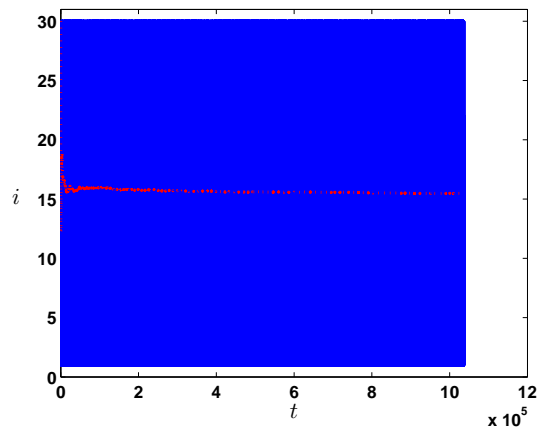}
\end{center}
\caption{
Same 
as  Fig. \protect{\ref{figa03}}  however now   
$\psi(\tau) =  5 \tau^{- 6}$ for $\tau > 1$ so $\alpha=5$.
After a short transient the time average
converges to the ensemble average on $15.5$ indicating ergodicity. 
}
\label{figa5}
\end{figure}

 In Figs. \ref{figa03},\ref{figa75} we show single particle trajectories
and their time average for $\alpha=0.3$ and $\alpha=0.75$ respectively.
The time average is clearly a fluctuating random variable, due to long
sticking times in states of the system. Notice that the particle visits
all lattice points, and the phase space is not decomposable into
inaccessible regions as found for strong ergodicity breaking. 
For a waiting time PDF with $\alpha=5$, we have a finite average waiting
time, and hence as shown in  Fig. \ref{figa5} 
we  find an ergodic behavior
$\overline{x}\simeq \langle x \rangle=(L+1)/2=15.5$ 
in the long time limit. 

\begin{figure}
\begin{center}
\epsfxsize=80mm
%\epsfbox{articlefigure3.eps}
\epsfbox{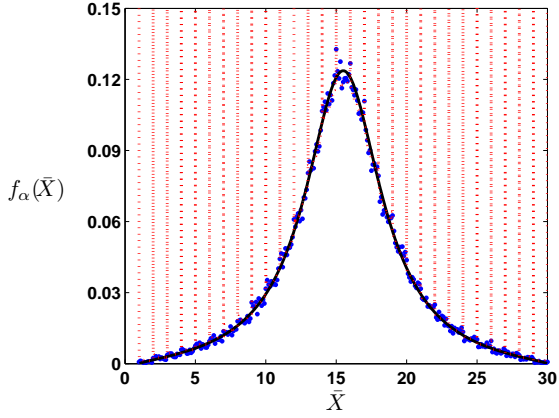}
\end{center}
\caption{
Distribution of the time average $\overline{x}$ 
for  
$\alpha = 3/4$.
Numerical simulations of the CTRW process
on a lattice (dots) are well approximated with
the continuum limit Eq.  
(\protect\ref{eq37}) (solid curve). 
The dotted curve with  divergences on the
lattice points is the analytical theory Eq. 
(\protect\ref{eq11}).
}
\label{fig4}
\end{figure}

\begin{figure}
\begin{center}
\epsfxsize=80mm
\epsfbox{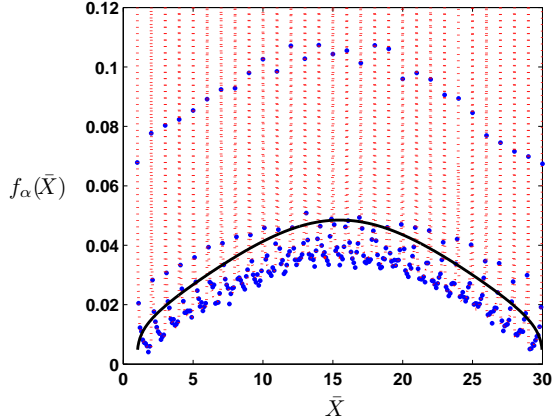}
\end{center}
\caption{
Same as Fig. \protect\ref{fig4} for $\alpha =0.3$.
Now the fluctuations are larger compared with the $\alpha=3/4$
case and the underlying structure of the lattice is important. Ofcourse
bin size must be made small enough for the lattice effect to be observed. 
}
\label{fig5}
\end{figure}

In Figs. \ref{fig4},\ref{fig5} we present
the PDF of $\overline{x}$ for $\alpha= 0.75$ and $\alpha = 0.3$ 
respectively.
 Comparison between simulations and
theory Eq.
(\ref{eq11}) show excellent agreement 
without fitting. 
In Eq. (\ref{eq11}) we use $p^{{\rm eq }} _i=1/L$ which is the obvious
population probability. 
The number of realizations
was $120000$ and the simulation
time $t=10^8,10^{12}$ for Figs. \ref{fig4},\ref{fig5}
respectively.
In Figs. \ref{fig4},\ref{fig5} we also show the continuum approximation
Eq. 
(\ref{eq37}). From Figs. \ref{fig4},\ref{fig5}
we see that the structure of the lattice
is encoded in the distribution of the time average $\overline{x}$.
Since the observable at any given moment of time attains the
values $x=1,...30$ we have $30$ divergences in Figs. \ref{fig4},\ref{fig5}
in agreement with the more general rule discussed under Eq.
(\ref{eq38ADIa}).
When the system is made large these effects become negligible and
the continuum approximation works well. 

\subsection{Non identical waiting time distributions}

 The CTRW considers a situation where a single waiting time
PDF $\psi(\tau)$ describes the dynamics. What happens when
 the waiting
times in the states $i$ are not identically distributed?  
For example consider the mentioned blinking quantum dots  
\cite{Dahan,Marg,Marg1,Marg3}.
The quantum dot when interacting
with a laser field will switch between an on state $(+)$
where many photons are emitted and an off $(-)$ state.  
The sojourn times in state on (and off) are independent identically
distributed random variables \cite{Bianco}. 
The PDF of sojourn times in states on and off follow power law
statistics 
atleast within a long
experiment time \cite{Marg3}
and in Laplace space
$\hat{\psi}_{\pm}(s) \sim 1 - A_{\pm} s^\alpha + \cdots$. 
This two state renewal process is characterized with 
two amplitudes $A_{+}$ and $A_{-}$
and in this sense it differs from the usual CTRW. 
It is worthwhile noting that
the visitation fraction in  states $\pm$ clearly satisfy
\begin{equation}
\lim_{N \to \infty} v_{\pm} = \lim_{N \to \infty} { n_{+} \over N}=\lim_{N \to \infty} {n_{-} \over N} = {1 \over 2}
\label{eq52}
\end{equation}
in the limit of long measurement time, where $N$ is the total number
of transitions between states on and off. 
If we consider an ensemble of $M$ independent  blinking dots
the population fraction of the number of dots in state on (+) $M_{+}$ and 
off (-) $M_{-}$ is 
\begin{equation}
p^{eq}_{\pm} = \lim_{M \to \infty} {M_{\pm} \over M} =
{A_{\pm} \over A_{-} + A_{+} }, 
\label{eq53}
\end{equation}
where the population fraction is measured in the limit of long
measurement time. 
So here the visitation fraction and the population fraction are
non-identical if $A_{+} \neq A_{-}$. 

 More generally we consider the renewal dynamics with power law waiting
times in each state however now
\begin{equation}
\hat{\psi}_i(s) \sim 1 - A_i s^\alpha
\label{eq54}
\end{equation}
when the Laplace variable  $s \to 0$, $A_i > 0$ for $i=1, \cdots,L$
and $0<\alpha<1$.
In this case the population fractions are 
related to the visitation fractions according to
\begin{equation}
p^{{\rm eq}} _i = \lim_{N \to \infty }  { A_i v_i \over \sum_{i =1} ^{L}  A_i v_i }
\label{eq55}
\end{equation}
and $w \cdot v = v$. 
Now the main Eqs. derived in this Sec. must be modified, for example  
Eq. (\ref{eq48})
\begin{equation}
\hat{f}_{s,\vec{{\cal O}} , N,  \vec{n}} \left( u \right) = {\Pi_{i=1}^L \hat{\psi}_{i} ^{n_i} \left( s + u {\cal O}_i \right) \left[ 1 - \hat{\psi}_i \left( s + u {\cal O}_k \right) \right] \over s + u {\cal O}_k }. 
\label{eq56}
\end{equation}
Then using Eqs. (\ref{eq54},\ref{eq55},\ref{eq56})
we find Eq. 
(\ref{eq11}) (the method is nearly identical to the case where
all the waiting times are identical). 
Thus while the dynamics clearly differs from the
usual CTRW with a single waiting time PDF, and the visitation
fraction is not identical to the population fraction, our main Eq.
for the distribution of the time averages Eq. (\ref{eq11})
is still valid.

 Another situation is when the system has different types of waiting
times, for example some states may have exponential waiting
times while others follow power law statistics.  Or we may
have some states with a power law waiting time PDF with an exponent
$0<\alpha_1<1$ and for other states an exponent 
$0<\alpha_2 <\alpha_1$.  
Also in this case our main result Eq. (\ref{eq11}) will be valid.
 In the long time limit the
system will occupy only the states with the smallest $\alpha<1$.
Only those states are relevant for the calculation of the distribution
of time averages Eq. (\ref{eq11}). Other states might be visited many times
so their visitation fraction is not necessarily small, still the
time the system spends in these states is short and they do
not contribute to the time average.     

\section{Discussion}

 To summarize we have obtained  very general
distributions of time averages
of physical observables 
of weakly non-ergodic systems Eqs. (\ref{eq11},\ref{eq32}).
Unlike usual ergodic statistical mechanics
where the time averages are equal to the ensemble averages,
we find large fluctuations of time averages.
Unlike strong ergodicity breaking, the space is not
separated into inaccessible regions.  Instead the
system explores all states and the number of visits per state $i$
$n_i$ is large.  However due to the power law sojourn
times the dynamics is non-stationary and non-ergodic. Since 
exploration of  space is possible in weak ergodicity breaking,
we were able to construct
a rather general statistical theory for the distribution of time averages
which is valid in the long time limit and does not depend on the
initial conditions of the system. 
Further, the exploration of the cells $i=1,...L$ 
 leads to relations
 between
the non-ergodic statistical properties of a single system,
 and the equilibrium populations
of
an ensemble of systems. Hence 
the distribution
of time averages 
Eq. (\ref{eq11}) 
depends on 
population probabilities $p^{{\rm eq}}_i$s
and when thermal detailed balance is applied these probabilities are
given by
Boltzmann's canonical law. 
Such behavior was found in two classes of models: (a) for systems
with a common power law waiting time PDF $\psi(\tau)$
 where the visitation fraction is
equal to the population fraction and (b) for systems with
different waiting times PDFs where the visitation fraction is 
not equal to the population fraction. In both cases the
population fraction is not identical to the occupation fraction,
the latter remaining a random variable when $0<\alpha<1$.
Due to the large number of applications of the CTRW  model, 
our theory may find its applications
in several systems. The mathematical foundation of the theory
is Lamperti's generalized arcsine law and L\'evy's statistics, while
usual type of statistical mechanics is based on the Gaussian
central limit theorem. Due to the deep relations between the stochastic
CTRW model and
other models of anomalous diffusion, e.g. the quenched trap model 
and
deterministic dynamics 
our non-ergodic theory might find more general justification.

{\bf Acknowledgment:} This work was supported by the
Israel Science Foundation. EB thanks G. Margolin, S. Burov,
and G. Bel for discussions.  

% Adi's comment:
%For the harmonic oscillator $V(x)=x^2$ the PDF at $\bar{x}=0$ is given by :
%$$ f_{\alpha}\left(\overline{{\cal 0}}=0\right)=\frac{ \Gamma(\frac{\alpha}{2}) \tan(\frac{\pi\alpha}{2})
%}{\Gamma(\frac{1+\alpha}{2}) \pi \sqrt{T} }=\frac{ 1}{
%\cos(\frac{\pi\alpha}{2})\Gamma(1-\frac{\alpha}{2}) \Gamma(\frac{1+\alpha}{2})
%\sqrt{T} }$$


\begin{thebibliography}{99}
\bibitem{remark}
See \cite{Feller,Metzler,Taqqu,Levy} and Ref. therein for properties
of one sided L\'evy stable laws. Note that 
the Laplace variable $u$
in Eq. (\ref{eq02})  is dimensionless. 
We may use
$\exp(- B p^{{\rm eq}}_i u^\alpha)$ in Eq. (\ref{eq02}) with 
$B>0$ however our final result
Eq.
(\ref{eq11})
does not depend on it so we take $B=1$.
\end{thebibliography}
\end{document}